\documentclass[conference,a4paper]{IEEEtran}

\usepackage{amsmath}
\usepackage{stackengine}

\usepackage{bm}
\usepackage{array}
\usepackage{tabu}

\usepackage{hyperref} 
\usepackage{graphics} 
\usepackage{algorithmic} 
\usepackage{bm}
\usepackage{graphicx}
\usepackage{subfigure}
\usepackage{subfloat}
\usepackage{amsmath}
\usepackage{caption}
\usepackage{amsthm,amssymb,amsmath}
\usepackage{lmodern}
\usepackage{subfigure}
\DeclareMathAlphabet{\mathpzc}{OT1}{pzc}{m}{it}
\usepackage{mathtools}

\usepackage{stackengine}

\usepackage{amsmath}

\usepackage{graphicx}
\usepackage[font={small}]{caption}
\usepackage{ptext}

\usepackage{algorithm}
\usepackage{algorithmic}
\usepackage{float}
\usepackage[utf8]{inputenc} 
\usepackage[T1]{fontenc}
\usepackage{url}
\usepackage{ifthen}
\usepackage{cite}

\usepackage{color}
\usepackage[dvipsnames]{xcolor}

\begin{document}

\title{\LARGE{Successive Wyner-Ziv Coding for the Binary CEO Problem under Log-Loss}}
\author{Mahdi Nangir$^1$, Reza Asvadi$^2$, Mahmoud Ahmadian-Attari$^1$, and Jun Chen$^3$\\
        \small{$^1$Faculty of Electrical Engineering, K.N. Toosi University of Technology, Tehran, Iran.} \\ 
        \small{$^2$Faculty of Electrical Engineering, Shahid Beheshti University, Tehran, Iran.} \\ \small{$^3$Department of Electrical and Computer Engineering, McMaster University, Hamilton, ON, Canada.}\\ 
        \small{Emails: mahdinangir@ee.kntu.ac.ir, r\_asvadi@sbu.ac.ir, mahmoud@eetd.kntu.ac.ir, junchen@ece.mcmaster.ca}}

\maketitle

\begin{abstract}
An $l$-link binary CEO problem is considered in this paper. We present a practical encoding and decoding scheme for this problem employing the graph-based codes. A successive coding scheme is proposed for converting an $l$-link binary CEO problem to the $(2l-1)$ single binary Wyner-Ziv (WZ) problems. By using the compound LDGM-LDPC codes, the theoretical bound of each binary WZ is asymptotically achievable. Our proposed decoder successively decodes the received data by employing the well-known Sum-Product (SP) algorithm and leverages them to reconstruct the source. The sum-rate distortion performance of our proposed coding scheme is compared with the theoretical bounds under the logarithmic loss (log-loss) criterion.
\end{abstract}

\section{Introduction}

Multi-terminal lossy source coding problems are applicable in the cooperative communications and distributed storage systems. In this paper, we specifically focus on the Chief Executive Officer (CEO) problem which appears in the wireless sensor networks (WSN). In a WSN, a target phenomenon is measured by independent sensors in a noisy environment. Then, the noisy observations are processed by some agents for being sent via independent links to a joint fusion center. In the coding literature, agents are called the encoders and the fusion center is called the joint CEO decoder. In digital applications, a binary symmetric source (BSS) is corrupted by independent binary noises which are modeled by the Bernoulli distribution. This scenario is called the binary CEO problem which has received less attention. In contrast, the quadratic Gaussian CEO problem where the target source and noises are Gaussian random processes is widely studied. Theoretical rate-distortion bounds are provided for the quadratic Gaussian CEO problem in \cite{oha98,PTR04,oha12}. Furthermore, some efficient coding and optimal rate-allocation schemes for the quadratic Gaussian CEO problem are presented in \cite{CZB04,CB08,BS09}. A successive WZ coding structure is given to achieve any point in the achievable rate-distortion region of the quadratic Gaussian CEO problem by applying the Berger-Tung coding that is optimal for this problem according to \cite{CB08}.

Source and channel codes with a graphical representation are powerful and practical tools for achieving theoretical bounds. They are implemented with iterative message-passing algorithms. Most of the network information theoretic problems, like the CEO problem, can be reduced to several point-to-point scenarios. Hence, an efficient design of the source and the channel codes leads to a good performance in the original multi-terminal coding problem. 
Nested LDPC and LDGM codes form a structure which efficiently performs close to the rate-distortion bound of the binary WZ problem \cite{WM09,NAA17}. For instance, consider a source code with the rate-distortion relation $R_s=1-h_b(d)+\epsilon_s$ for a target distortion $d$ and a channel code of rate $R_c=1-h_b(p*d)-\epsilon_c$ are utilized in the compound coding scheme for sufficiently small positive values $\epsilon_s$ and $\epsilon_c$. Let also $p*d=p(1-d)+d(1-p)$ and $p$ be the cross-over probability of the binary symmetric channel (BSC) between the source and the side information available at the decoder. In this case, it is shown that the binary WZ theoretical bound can be attainable \cite{WM09}.

An extension of the WZ problem can be considered in the multi-link scenarios like the CEO problem. Researches on the binary CEO problem are in two categories. First, works that focus on finding some tight bounds of the rate-distortion region, that is still an open problem in general. Only in the case of the log-loss criterion as a distortion measure, the exact rate-distortion bound is obtained in \cite{CW14}. Second, works which devise practical and low-complexity coding schemes to achieve the existing theoretical bounds, e.g., \cite{YSXZ08}, \cite{CB08}, and \cite{BS07}. Most of the researches in these categories are on the quadratic Gaussian CEO problem. Alternatively, our main object is to design an efficient coding scheme for the binary CEO problem that lies in the second category.

The main contribution of this paper is to devise a practical coding scheme for the binary CEO problem based on the successive WZ coding scheme and the source splitting idea \cite{RU97}, with the sum-rate distortion performance close to the theoretical bounds under the log-loss. By using the source splitting method, an arbitrary point in the achievable rate-distortion region of an $l$-link case is converted to a corner point in the achievable rate-distortion region of an $(2l-1)$-link case \cite{RU97}. In our scheme, the Bias-Propagation (BiP) \cite{FILL07, FF07} and the SP \cite{LXG02} algorithms are applied for the binary quantization and the syndrome-decoding, respectively.

\section{Problem Statement and System Model}

Let $X=(x_1,x_2,\cdots,x_n)$ be a remote BSS. Let $l$ noisy observations of $X$ be available through $l$ links that are jointly independent from each other. The noises of links are modeled by independent Bernoulli random variables $N_i$ with parameters $p_i$ for $1 \le i \le l$. The noisy observations are denoted by $Y_i=(y_{i,1},y_{i,2},\cdots,y_{i,n})$. In each link, an encoder compresses the noisy observation to a codeword denoted by $C_i$. 
The codewords $C_i$ are sent to the CEO decoder via noiseless channels. Finally, the CEO decoder produces the reconstruction sequence $\hat{X}=(\hat{x}_1,\hat{x}_2,\cdots,\hat{x}_n)$.
We consider the soft reconstruction with $\hat{x}_j$ being an estimate of the a posteriori distribution of $x_j$ given $C_1,\cdots,C_l$. Based on the definitions in \cite{CW14}, the symbol-wise and the total log-loss are respectively as follows:
\begin{align}
\label{log-loss}
d(x_j,\hat x_j)&=\log_2 \big({1 \over \hat x_j(x_j)}\big), \ \ \ j=1,\cdots,n, \\ \nonumber
D_{\text{em}}&={1 \over n} \sum_{j=1}^n \log_2 \big({1 \over \hat x_j(x_j)}\big).
\end{align}

\begin{figure}[t!]
	\begin{center}
		\centering
		\includegraphics[width=2.6in,height=1.6in]{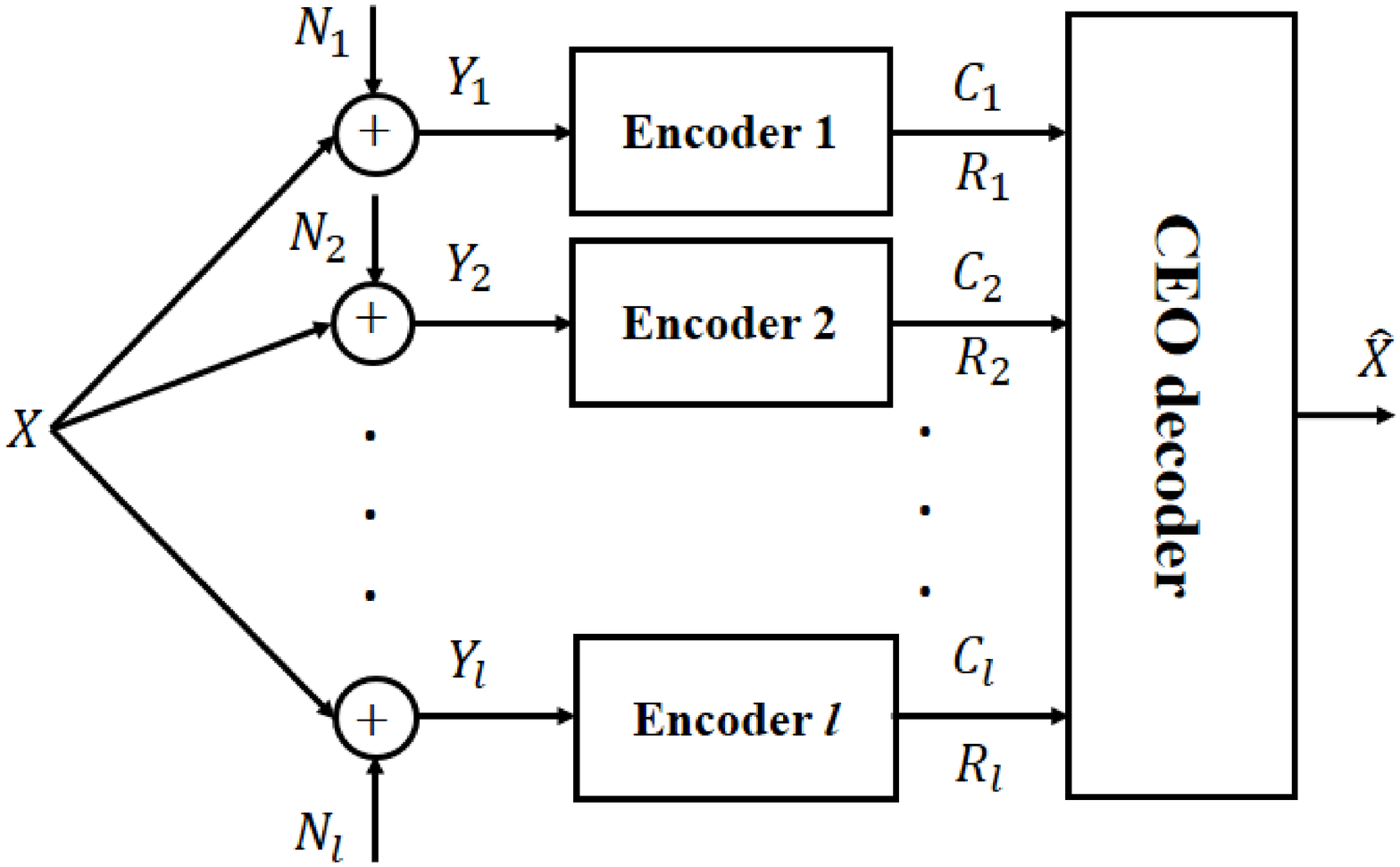}
	\end{center}
	\vspace{-10pt}
	\caption{Configuration of an $l$-link binary CEO problem.}
	\label{CEO}
\end{figure}

The block diagram of an $l$-link binary CEO problem is depicted in the Fig. \ref{CEO}. Assume links are rate-limited, i.e., the $i$-th link can communicate data at most with rate $R_i$ for $1 \le i \le l$.

We suppose that the encoders cannot collaborate with each other, hence $R_i$s are independently determined. In the context of the distributed lossy source coding problems, collaboration of the encoders affects the achievable rate-distortion region. The Berger-Tung inner bound under the log-loss is tight \cite{CW14}, and it is mentioned in the following definition.

\textit{Definition 1:} An $l$-tuple of rates $(R_1,R_2,\cdots,R_l)$ is achievable with distortion $D$ for the binary CEO problem in Fig. \ref{CEO}, if for any nonempty set $\mathcal{A} \subseteq \{1,2,\cdots,l\}$,
\begin{align}
\label{BTinner}
\sum_{i \in \mathcal{A}} R_i &\ge I(Y_{\mathcal{A}};U_{\mathcal{A}} | U_{\mathcal{A}^c},Q), \\ \nonumber
D &\ge H(X | U_1,\cdots,U_l,Q),
\end{align}
for some conditional pmf $p(q)p(u_1|y_1,q) \cdots p(u_l|y_l,q)$, where $U_i$s are the quantized sequences and $Q$ is an auxiliary timesharing random variable with a sample realized by $q$. In (\ref{BTinner}), $Y_{\mathcal{A}}=\{Y_i:i \in \mathcal{A}\}$ and $\mathcal{A}^c=\{1,2,\cdots,l\}-\mathcal{A}$.

The following bounds on the alphabet sizes of $U_i$ and $Q$ in (\ref{BTinner}) are provided \cite{CW14},
\begin{align}
\label{cardin}
|\mathcal{U}_i| &\le |\mathcal{Y}_i|, \ \ \ \ \text{for}  \ \ \ \  1 \le i \le l, \\ \nonumber
|\mathcal{Q}| &\le l+2.
\end{align}

In this paper, we consider the log-loss criterion and evaluate the performance of the proposed coding scheme by measuring the gap values from (\ref{BTinner}).
Furthermore, we assume BSC test-channel models for the encoders in each link, i.e., the Encoder $i$ is modeled by a BSC with the crossover probability $d_i$. Note that as far as the sum-rate distortion function is concerned, the timesharing variable $Q$ can be assumed to be a constant and hence it is eliminated \cite{NAAC18}. Under these assumptions, the sum-rate and distortion bounds are derived,
\begin{align}
\label{eq1}
&I({U}_1,\cdots,{U}_l;{Y}_1,\cdots,{Y}_l) = H(U_1,\cdots,U_l)-\sum_{i=1}^l h_b(d_i), \\ \nonumber
&H({X}|{U}_1,\cdots,{U}_l) =1+\sum_{i=1}^l h_b(P_i)-H(U_1,\cdots,U_l),
\end{align}
where $P_i=d_i*p_i$, $h_b(x)=-x\log_2x-(1-x)\log_2(1-x)$, and
\begin{align}
\label{eq2}
H(U_1,\cdots,U_l)=-\sum_{j=1}^{2^{l-1}} [\nu_j+\nu_{{2^l+1}-j}] \log_2 [{\nu_j+\nu_{{2^l+1}-j} \over 2}],
\end{align}
such that
\begin{align}
\label{eq3}
\nu_j&=\Pr\{U_1=u_1,\cdots,U_l=u_l | X=0\} \\ \nonumber
&=\Pr\{U_1=\bar{u_1},\cdots,U_l=\bar{u_l} | X=1\}, \ \ \text{for} \ \ j=1,\cdots,2^l.
\end{align}
In (\ref{eq3}), $u_1 u_2 \cdots u_l$ is the binary conversion of the decimal number $j-1$ in $l$ bits, $\bar{0}=1$, and $\bar{1}=0$.

An interesting optimization problem arises from a general $l$-link binary CEO problem with BSC test-channels. This problem is finding the optimal values of $d_i$ for a specified noise parameters $p_i$, $1 \le i \le l$. In this case, the optimality means the minimization of the sum-rate bound subject to a fixed value of distortion bound or vice versa, i.e.,
\begin{align}
\label{opt1}
\underset{d_1,\cdots,d_l}{\text{min}}& \ \ I({U}_1,\cdots,{U}_l;{Y}_1,\cdots,{Y}_l), \\ \nonumber
\text{s.t.}& \ \  H({X}|{U}_1,\cdots,{U}_l)= D_{\text{log-loss}}.
\end{align}
An unconstrained form of (\ref{opt1}) by using the Lagrangian multiplier $\mu$ can be written as follows:
\begin{align}
\label{opt2}
\underset{d_1,\cdots,d_l}{\text{min}} \ \ H({X}|{U}_1,\cdots,{U}_l) + {\mu} I({U}_1,\cdots,{U}_l;{Y}_1,\cdots,{Y}_l).
\end{align}
Since the problem (\ref{opt2}) is not convex \cite{NAAC18}, it can be numerically solved by an $l$-dimensional exhaustive search in the set $[0,0.5]^l$. 
In Section IV, these theoretical bounds are presented for some cases.

\section{The Encoding and Decoding Schemes}

In this section, we present the encoding and decoding procedures. Our proposed coding scheme is practical and it is implemented by employing graph-based coding approach.

\subsection{The Proposed Encoding Scheme}
\begin{figure}[t!]
	\begin{center}
		\centering
		\includegraphics[width=3in,height=2in]{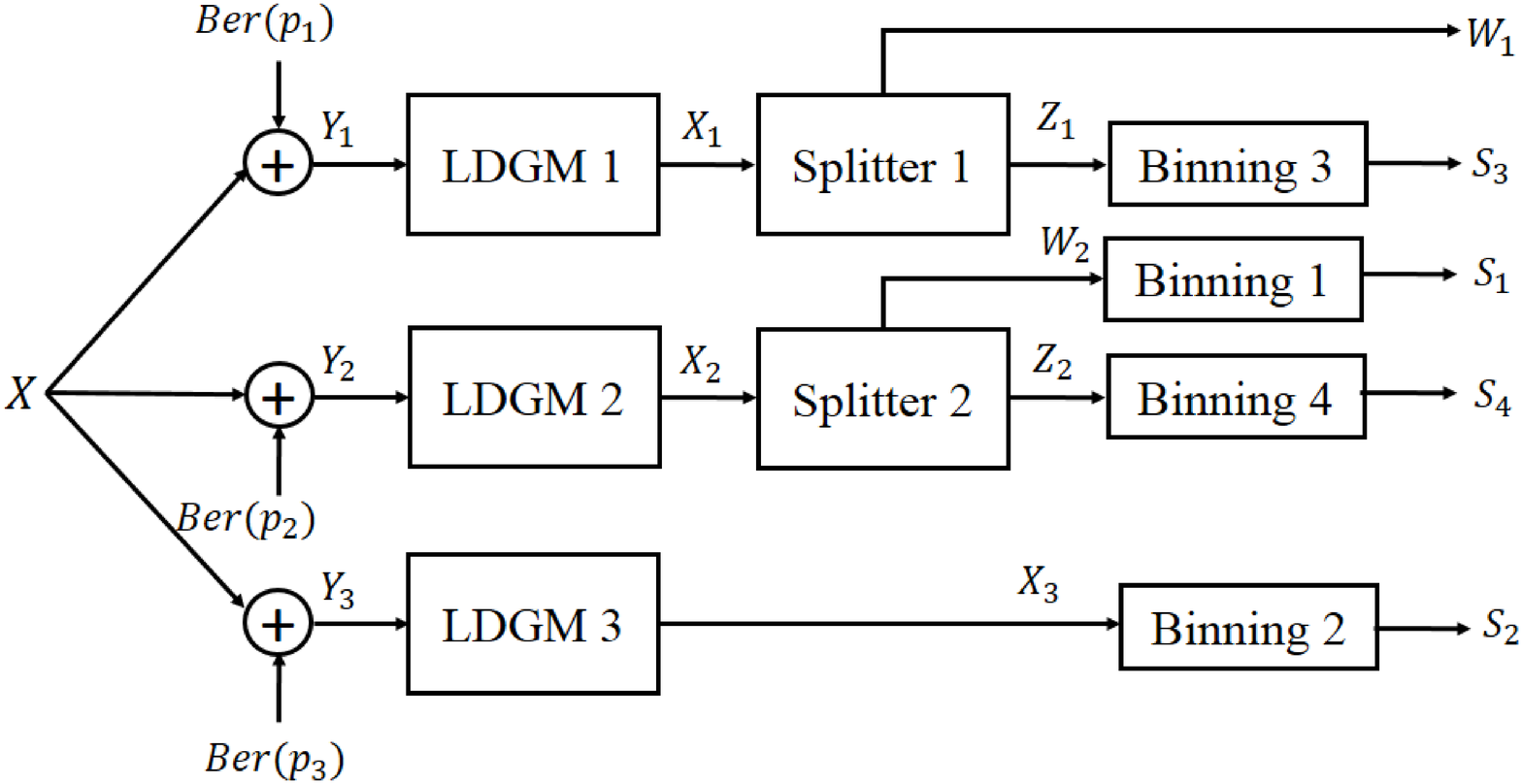}
	\end{center}
	\vspace{-10pt}
	\caption{The proposed encoding scheme}
	\label{ENC}
\end{figure}

The encoding procedure in each link includes three steps: (i) Quantization, (ii) Splitting, and (iii) Binning. In the following, we illustrate their details for the case of $l=3$ as it is exhibited in Fig. \ref{ENC}. This scheme can be easily generalized for $l>3$.

In the first step, the noise corrupted observations $Y_i$ are quantized to LDGM codeword $X_i$; $i=1,2,3$. In this step, a BiP algorithm is applied in the $i$-th LDGM. Assume the Hamming distortion of the quantization step is $d_i$. It is shown in \cite{FF07} that the binary quantization using LDGM codes have a performance near the rate-distortion limit for a BSS, i.e., the $i$-th LDGM code rate can approach to $1-h_b(d_i)$. We utilize the LDGM part of a compound LDGM-LDPC code for accomplishing the first step of the encoding process.

In the second step, the source splitting is fulfilled in the first and the second links. Thereby, five sequences $\{W_1,Z_1,W_2,Z_2,X_3\}$ are the outputs of the splitting step, which are used for the syndrome generation in the next step.

In the third step, we use a binning method to generate the syndromes. Actually, this step is known as Slepian-Wolf coding, which is accomplished by employing channel codes. In our proposed coding scheme, the binning is realized by the LDPC code of the compound structures. $W_1$ is obtained without any binning and the binning blocks $1$ to $4$ generate syndromes $S_1$ to $S_4$, respectively. Finally, $\{W_1,S_1,S_2,S_3,S_4\}$ are sent to the CEO decoder.

\subsection{The Proposed Decoding Scheme}

\begin{figure}[t!]
	\begin{center}
		\centering
		\includegraphics[width=2.5in,height=2.2in]{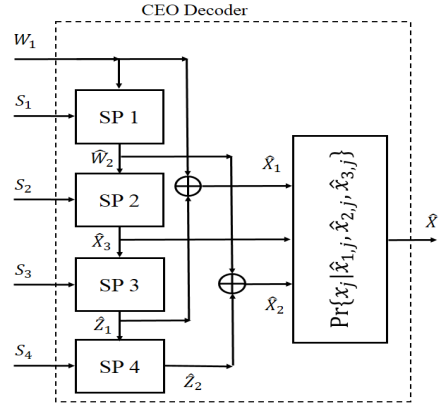}
	\end{center}
	\vspace{-10pt}
	\caption{The proposed decoding scheme}
	\label{DEC}
\end{figure}

In our proposed decoding algorithm, we exploit the syndrome-based decoding method by utilizing LDPC codes. The decoding scheme contains two steps. The first one is reconstruction of the quantized codewords $X_1$, $X_2$, and $X_3$. This step is executed by using the SP algorithm, successively. The second one is a soft decision rule which produces the reconstruction $\hat X$. The block diagram of the decoder is illustrated in Fig. \ref{DEC}.

If the LDPC codes applied in the SP algorithm designed with appropriate rates and degree distributions, then $\{W_2,X_3,Z_1,Z_2\}$ are decoded with a low bit error rate (BER). Hence, $\{X_1,X_2,X_3\}$ are decoded with low BER.

\subsection{A Practical Analysis of the Proposed Coding Scheme}
In this subsection, a heuristic analysis for the proposed coding scheme is provided. This analysis establishes some rate-distortion-block length relations for a $3$-link binary CEO problem. All of the numerical results in the next section can be interpreted based on this analysis. Let the length of the sequences be $n$. Presume size of the generator matrix of the LDGM code used in the binary quantization part is $m_i \times n$ for $i=1,2,3$. Thereby, for the distortion $d_i$:
\begin{align}
\label{e1}
{m_i \over n}=1-h_b(d_i)+\epsilon_i, \ \ \ \ \ \ \text{for} \ \ \ \ \ i=1,2,3,
\end{align}
where $\epsilon_i > 0$. By properly designing the LDGM codes and increasing the length $n$, one can achieve a small enough value of $\epsilon_i$.

In the syndrome-decoding part, which is done by the successive SP algorithms, if the optimized degree distributions are considered for the BSC and sufficiently long-length LDPC codes are used, then the BER in reconstruction of $\{X_1,X_2,X_3\}$ can be made very close to zero, i.e., $\text{BER}_i \to 0$. In such a case, the total Hamming distortion of the $i$-th link approximately equals $d_i$.

Each splitter in the proposed scheme has one input (say $X_i$) and two outputs (say $W_i$ and $Z_i$). Clearly, there exists a Hamming distortion between $X_i$ and $W_i$ (similarly between $X_i$ and $Z_i$), that might be considerable. It is important to note that the distortion between $X_i$ and the pair $(W_i,Z_i)$ is zero, due to a one-to-one correspondence between them. Hence, the reconstruction of $W_i$ or $Z_i$ is not the final target by itself and the source splitters do not affect $d_i$ in each link. Thus, the following Hamming distortions are defined for the splitters,
\begin{align}
\label{diss}
\alpha_i=\mathbb{E} \{{1 \over n} \sum_{j=1}^n [y_{i,j} \oplus w_{i,j}]\}, \ \ \ \beta_i=\mathbb{E} \{{1 \over n} \sum_{j=1}^n [y_{i,j} \oplus z_{i,j}]\},
\end{align}
where $\oplus$ shows the binary summation.
Suppose that the rate of $i$-th LDPC code utilized in the $i$-th binning block and the $i$-th SP algorithm is as follows, for $i=1,2,3,4$:

\begin{align}
\label{LDPC}
\text{LDPC 1}&: \ \ {m_2-k_1 \over n}, \ \ \ \ \ \text{LDPC 2}: \ \ {m_3-k_2 \over n}, \\ \nonumber 
\text{LDPC 3}&: \ \ {m_1-k_3 \over n}, \ \ \ \ \ \text{LDPC 4}: \ \ {m_2-k_4 \over n}.
\end{align}
Hence, four compound LDGM-LDPC codes are engaged in the proposed coding scheme. They respectively employ LDGM codes $1$,$2$,$2$, and $3$ with their associated LDPC codes $3$,$1$,$4$, and $2$. The rate-distortion relations in the LDPC codes are taken as follows:
\begin{align}
\label{LDPCrates}
{m_2-k_1 \over n}&=1-h_b(p_1*p_2*\alpha_1*\alpha_2)-\delta_1, \\ \nonumber 
{m_3-k_2 \over n}&=1-h_b(p_2*P_3*\alpha_2)-\delta_2, \\ \nonumber 
{m_1-k_3 \over n}&=1-h_b(p_1*P_3*\beta_1)-\delta_3, \\ \nonumber 
{m_2-k_4 \over n}&=1-h_b(p_1*p_2*\beta_1*\beta_2)-\delta_4.
\end{align}
In (\ref{LDPCrates}), $\delta_i>0$ for $i=1,2,3,4$, which can be made sufficiently close to zero through appropriate code design.

\section{Results and performance analysis}

In this section, we present some numerical and analytical results to demonstrate performance of the proposed coding scheme for some cases. The employed degree distributions are optimized over BSC using the density evolution technique \cite{SC07}. Furthermore, the degree distributions of the LDGM codes are designed using the method described in \cite{NAA17}.
The following source splitting rules are utilized in our implementations:
\begin{align}
	\label{split}
	W_j&=(x_{j,1},x_{j,2},\cdots,x_{j,n'}), \\ \nonumber
	Z_j&=(x_{j,n'+1},x_{j,n'+2},\cdots,x_{j,n}),
\end{align}
where $1 \le n' \le n$ and $j=1,2$. Clearly in (\ref{split}), $X_j=W_jZ_j$. Another is that
\begin{align}
	\label{splitt}
	W_j &= \Psi_n^{-1} \bigg[ \min \big(\Psi_n(X_j),T_j \big) \bigg], \\ \nonumber 
	Z_j=& \Psi_n^{-1} \bigg[ \max \big(\Psi_n(X_j),T_j \big) - T_j \bigg] ,
\end{align}
where $T_j \in \{0,1,\cdots, 2^{n}-1 \}$ for $j=1,2$, and $\Psi_n$ is the $n$-tuple binary to decimal conversion function. In (\ref{splitt}), we have
\begin{align}
	\label{sum}
	{X_j} = W_j \oplus Z_j.
\end{align}
Therefore, these rules yield a one-to-one correspondence between the input and the outputs of the splitters.

\begin{figure}[t!]
	\begin{center}
		\centering
		\includegraphics[width=3.3in,height=2.2in]{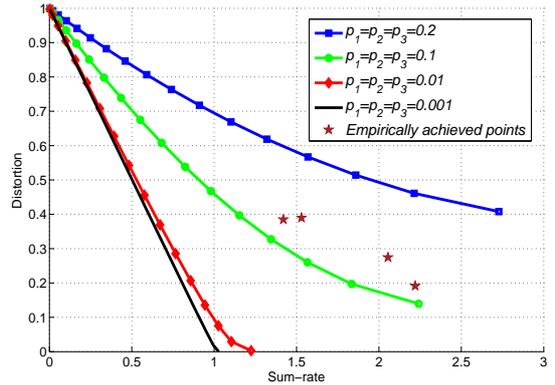}
	\end{center}
	\vspace{-10pt}
	\caption{\small{The sum-rate distortion bound in the case of equal noise parameters with a same quantizer to each link.}}
	\label{fig1}
\end{figure}

\textit{Example 1:} Consider equal observation noise parameters, i.e., $p_1=p_2=p_3$. We assume $d_1=d_2=d_3$, i.e., a same LDGM code quantizes each observation. By this assumption, the sum-rate distortion curve is obtained from (\ref{eq1}) as depicted in Fig. \ref{fig1} for different noise parameters. The parameters used in our coding scheme are given in Table \ref{t2}. We have assumed the code length $n=10000$ and the noise parameters $p_1=p_2=p_3=0.1$. Two examples for the corner and the intermediate points are provided in Table \ref{t2}. We have used the source splitting rule (\ref{split}) with $n'={n \over 2}$, for implementing codes that achieve the intermediate points. Hence,
\begin{align}
\label{rs}
R_1={m_1+k_3 \over {2n}}, \ \ R_2={k_1+k_4 \over {2n}}, \ \ R_3={k_2 \over {n}}.
\end{align}
An improved performance can be achieved by increasing $n$, $l$, and also using better source splitters.

\begin{table*}[t!]
	\caption{\footnotesize{PARAMETERS AND NUMERICAL RESULTS OF THE PROPOSED ENCODING AND DECODING SCHEMES.}}
	\label{t2}
	\centering
	\vspace{-5pt}
	\begin{center}
		\scalebox{0.8}{
			\begin{tabular} {| c | c | c | c | c | c | c | c | c | c |}
				\hline 
				$m_i$ & $k_1,k_2,k_3,k_4$ & $\alpha_1,\alpha_2$ & $\beta_1,\beta_2$ & $d_1,d_2,d_3$ & $R_1,R_2,R_3$ & $\text{BER}_1$,$\text{BER}_2$,$\text{BER}_3$,$\text{BER}_4$ &  $D_{\text{log-loss}}$ & $D_{\text{em}}$ &  $\text{Gap}$ \\
				\hline
				$5400$ & $4400,4400,-,-$ & $d_1,d_2$ & $-,-$ & $0.102,0.1031,0.1025$ & $0.54,0.44,0.44$ & $0.0018$,$0.0025$,$-$,$-$ &  $0.3537$ & $0.385$ &  $0.0313$ \\
				\hline
				$9200$ & $6500,6500,-,-$ & $d_1,d_2$ & $-,-$ & $0.0137,0.0135,0.015$ & $0.92,0.65,0.65$ & $0.001$,$0.0016$,$-$,$-$ &  $0.1637$ & $0.1917$ &  $0.028$ \\
				\hline
				$5400$ & $5300,4800,5000,5300$ & $0.2785,0.253$ & $0.291,0.266$ & $0.102,0.1031,0.1025$ & $0.52,0.53,0.48$ & $0.0024$,$0.003$,$0.003$,$0.0038$ &  $0.3537$ & $0.3898$ &  $0.0362$ \\
				\hline
				$7200$ & $7100,6600,6500,7100$ & $0.269,0.2536$ & $0.2486,0.2521$ & $0.052,0.0533,0.0511$ & $0.685,0.71,0.66$ & $0.002$,$0.0021$,$0.0027$,$0.0031$ &  $0.241$ & $0.2747$ &  $0.0337$ \\
				\hline
		\end{tabular}}
	\end{center}
\end{table*}

Based on the numerical and analytical results in \cite{NAAC18} for a two-link binary CEO problem, the allocation $d_1=d_2$ is not optimal for some values of the sum-rate and distortion, even in the case of equal noise parameters $p_1=p_2$. This surprising result is also true for the multi-link case. In the next example, a $3$-link binary CEO problem is considered with almost prominent difference between the noise parameters.

\begin{figure}[t!]
	\begin{center}
		\centering
		\includegraphics[width=3.3in,height=2.2in]{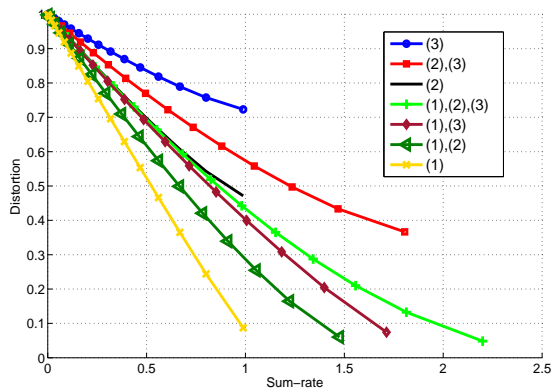}
	\end{center}
	\vspace{-10pt}
	\caption{\small{The sum-rate distortion bounds (Example $2$). The number of involved links are given in the legend.}}
	\label{fig4}
\end{figure}

\textit{Example 2:} Let $p_1=0.01$, $p_2=0.1$, and $p_3=0.2$. The sum-rate distortion bounds are presented in Fig. \ref{fig4}. It is assumed that the binary quantizers in each link are the same, when more than one link are involved in sending information. Clearly, involving all of the links does not necessarily give an optimal allocation scheme for the distortion parameters $d_1$, $d_2$, and $d_3$. 

\begin{figure}[t!]
	\begin{center}
		\centering
		\includegraphics[width=3.3in,height=2.2in]{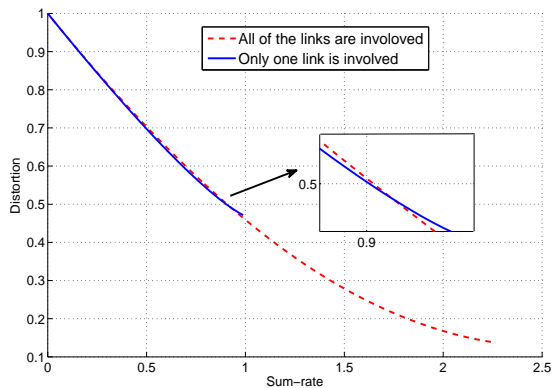}
	\end{center}
	\vspace{-10pt}
	\caption{\small{The sum-rate distortion bounds for $l=3$ and $p_1=p_2=p_3=0.1$.}}
	\label{fig3a}
\end{figure}

For more intuition, the sum-rate distortion bounds for $p_1=p_2=p_3=0.1$ is depicted in Fig. \ref{fig3a} for two cases of involving links. It is seen that a fair allocation of the parameters $d_1$, $d_2$, and $d_3$ does not provide the minimum sum-rate distortion bound, even with the equal noise parameters.

\section{Conclusion}

In this paper, an $l$-link binary CEO problem has been investigated under the log-loss. We also devised a practical coding scheme for this problem. By using the source splitting technique, an arbitrary point in the $l$-dimensional achievable rate-region has been converted to a corner point of an $(2l-1)$-dimensional rate region with applying a successive Wyner-Ziv encoding scheme. We presented a CEO decoder which decodes the observations successively and then it reconstructs the remote source outputs in a soft manner. Our numerical results confirmed that the performance of the proposed scheme is close to the existent theoretical bounds.

\bibliographystyle{IEEEtran}

\bibliography{refs}

\begin{thebibliography}{10}
\providecommand{\url}[1]{#1}
\csname url@samestyle\endcsname
\providecommand{\newblock}{\relax}
\providecommand{\bibinfo}[2]{#2}
\providecommand{\BIBentrySTDinterwordspacing}{\spaceskip=0pt\relax}
\providecommand{\BIBentryALTinterwordstretchfactor}{4}
\providecommand{\BIBentryALTinterwordspacing}{\spaceskip=\fontdimen2\font plus
\BIBentryALTinterwordstretchfactor\fontdimen3\font minus
  \fontdimen4\font\relax}
\providecommand{\BIBforeignlanguage}[2]{{%
\expandafter\ifx\csname l@#1\endcsname\relax
\typeout{** WARNING: IEEEtran.bst: No hyphenation pattern has been}%
\typeout{** loaded for the language `#1'. Using the pattern for}%
\typeout{** the default language instead.}%
\else
\language=\csname l@#1\endcsname
\fi
#2}}
\providecommand{\BIBdecl}{\relax}
\BIBdecl

\bibitem{oha98}
Y.~Oohama, ``The rate-distortion function for the quadratic {G}aussian {CEO}
  problem,'' \emph{IEEE Transactions on Information Theory}, vol.~44, no.~3,
  pp. 1057--1070, 1998.

\bibitem{PTR04}
V.~Prabhakaran, D.~Tse, and K.~Ramachandran, ``Rate region of the quadratic
  {G}aussian {CEO} problem,'' in \emph{Information Theory, 2004. Proceedings.
  International Symposium on}.\hskip 1em plus 0.5em minus 0.4em\relax IEEE,
  2004, p. 119.

\bibitem{oha12}
Y.~Oohama, ``Distributed source coding of correlated {G}aussian remote
  sources,'' \emph{IEEE Transactions on Information Theory}, vol.~58, no.~8,
  pp. 5059--5085, 2012.

\bibitem{CZB04}
J.~Chen, X.~Zhang, T.~Berger, and S.~B. Wicker, ``An upper bound on the
  sum-rate distortion function and its corresponding rate allocation schemes
  for the {CEO} problem,'' \emph{IEEE Journal on Selected Areas in
  Communications}, vol.~22, no.~6, pp. 977--987, 2004.

\bibitem{CB08}
J.~Chen and T.~Berger, ``Successive {W}yner-{Z}iv coding scheme and its
  application to the quadratic {G}aussian {CEO} problem,'' \emph{IEEE
  Transactions on Information Theory}, vol.~54, no.~4, pp. 1586--1603, 2008.

\bibitem{BS09}
H.~Behroozi and M.~R. Soleymani, ``Optimal rate allocation in successively
  structured {G}aussian {CEO} problem,'' \emph{IEEE Transactions on Wireless
  Communications}, vol.~8, no.~2, pp. 627--632, 2009.

\bibitem{WM09}
M.~J. Wainwright and E.~Martinian, ``Low-density graph codes that are optimal
  for binning and coding with side information,'' \emph{IEEE Transactions on
  Information Theory}, vol.~55, no.~3, pp. 1061--1079, March 2009.

\bibitem{NAA17}
M.~Nangir, M.~Ahmadian-Attari, and R.~Asvadi, ``Binary {W}yner–{Z}iv code
  design based on compound {LDGM–LDPC} structures,'' \emph{IET
  Communications}, vol.~12, pp. 375--383(8), March 2018.

\bibitem{CW14}
T.~A. Courtade and T.~Weissman, ``Multiterminal source coding under logarithmic
  loss,'' \emph{IEEE Transactions on Information Theory}, vol.~60, no.~1, pp.
  740--761, Jan 2014.

\bibitem{YSXZ08}
Y.~Yang, V.~Stankovic, Z.~Xiong, and W.~Zhao, ``On multiterminal source code
  design,'' \emph{IEEE Transactions on Information Theory}, vol.~54, no.~5, pp.
  2278--2302, May 2008.

\bibitem{BS07}
H.~Behroozi and M.~R. Soleymani, ``Distortion sum-rate performance of
  successive coding strategy in quadratic {G}aussian {CEO} problem,''
  \emph{IEEE Transactions on Wireless Communications}, vol.~6, no.~12, pp.
  4361--4365, December 2007.

\bibitem{RU97}
B.~Rimoldi and R.~Urbanke, ``Asynchronous {S}lepian-{W}olf coding via
  source-splitting,'' in \emph{Information Theory, 1997. Proceedings.
  International Symposium on}.\hskip 1em plus 0.5em minus 0.4em\relax IEEE,
  1997, p. 271.

\bibitem{FILL07}
T.~Filler, ``Minimizing embedding impact in steganography using low density
  codes,'' \emph{Master's thesis, SUNY Binghamton}, 2007.

\bibitem{FF07}
T.~Filler and J.~Fridrich, ``Binary quantization using belief propagation with
  decimation over factor graphs of {LDGM} codes,'' in \emph{Proc. Allerton
  Conf. Commun., Control, Comput.}\hskip 1em plus 0.5em minus 0.4em\relax IEEE,
  2007, pp. 1--5.

\bibitem{LXG02}
A.~D. Liveris, Z.~Xiong, and C.~N. Georghiades, ``Compression of binary sources
  with side information at the decoder using {LDPC} codes,'' \emph{IEEE
  communications letters}, vol.~6, no.~10, pp. 440--442, 2002.

\bibitem{NAAC18}
M.~Nangir, R.~Asvadi, M.~Ahmadian-Attari, and J.~Chen, ``Analysis and code
  design for the binary {CEO} problem under logarithmic loss,'' \emph{arXiv
  preprint arXiv:1801.00435}, 2018.

\bibitem{SC07}
D.~H. Schonberg, \emph{Practical distributed source coding and its application
  to the compression of encrypted data}.\hskip 1em plus 0.5em minus 0.4em\relax
  University of California, Berkeley, 2007.

\end{thebibliography}

\end{document}